\documentstyle[12pt]{article}
\begin{document}
\begin{flushright}
\today \\
IPM/P--99/080\\
\end{flushright}
\[ \]
\begin{center}
\section*{Junction equations for two spherically symmetric spacetimes
and the distributional method}
\[ \]
{\bf S. Khakshournia}

Sharif University of Technology, Department of Physics, Tehran, Iran.
\[\]
{\bf R. Mansouri}

Sharif University of Technology, Department of Physics, Tehran, Iran.

and

Institute for studies in Physics and Mathematics. P.O.Box 5531, Tehran, Iran.
\[ \]
\end{center}
\[ \]
\begin{center}
\subsection*{Abstract}
\end{center}
Applying the distributional formalism to study the dynamics of thin shells
in general relativity, we regain the junction equations for matching of two 
spherically symmetric spacetimes separated by a singular
hypersurface. In particular, we have shown how to define and insert the
relevant sign functions in the junction equations corresponding to the signs
of the extrinsic curvature tensor occurred in the Darmois--Israel method.
\newpage
\section{Introduction}

\hspace*{0.5cm}
Recently a distributional method has been developed to solve directly the
Einstein's field equations for thin shells embedded in an arbitrary 
space--time[1]. This method requires construction of space--time coordinates that 
match continuously on the shell and in which the four--metric based on the 
Lichnerowicz condition [2], is continuous, but has a finite jump in its 
first derivatives on the shell, so that its curvature tensor will contain 
a Dirac delta function. \\
So far the spherical thin shells in static spherically symmetric spacetimes
and the case of cylindrically symmetric thin layers have been solved by this
method[3,4]. In addition, an explicit formulation of distributional method
for handling nonlightlike surface layers and its equivalence to the jump
conditions of Darmois--Israel method through the analysis of the Bianchi
identities has been represented [1]. \\
On the other hand, it has been shown in the last years that in the framework
of Darmois--Israel formalism, the sign of the angular component of the
extrinsic curvature tensor of the shell, namely $K_{\theta}^{\theta}$, play a
key role in the classification of the global spherically symmetric space--time
structures, as mentioned by Sato [5], Berezin, Kuzmin, Tkachev [6] and
Sakai and Maeda [7]. This sign function gives us information related to
increase or decrease of the relevant coordinate in the direction normal to
the shell. In the distributional method of Mansouri--Khorrami mentioned above
this sign function, in spite of its importance, has not yet been introduced.
The aim of this paper is to show how one can generate the necessary sign
functions in the framework of Mansouri--Khorrami distributional method. In
this way we will show the full equivalence of the sign functions introduced
and their role in the junction equations with the corresponding signs of the
extrinsic curvature $K_{\theta}^{\theta}$ on the both sides of the shell in
the Darmois--Israel approach. \\
The paper is arranged as follows. In section 2 we review shortly the
distributional method for thin shells due to Mansouri and Khorrami. In
section 3 this distributional formalism is applied to the junction of two
Schwarzchild-de Sitter space--times through a timelike spherical thin
shell. Section 4 is devoted to the junction equations for two FRW space-times
bounded by a nonlightlike spherical thin shell.\\
{\it Conventions and definitions}: We use the signature ($+ - - -$) and put
$c=1$. Greek indices run from 0 to 3. Overdot denotes differentiation with
respect to the shell proper time $\tau$. The square brackets, [F], are used
to indicate the jump of any quantity $F$ across the shell. As we are going to
work with distributional valued tensors, there may be terms in a tensor
quantity $F$ proportional to some $\delta$--function. These terms are
indicated by $\check{F}$.

\section{Distributional Method}

\hspace*{0.5cm}Consider a space--time manifold $M$ consisting of overlapping
domains $M_{+}$ and $M_{-}$ with metrics $g_{\alpha \beta}^{+} (x_{+}^{\mu})$
and $g_{\alpha \beta}^{-}(x_{-}^{\mu})$ in terms of independent disconnected
charts $x_{+}^{\mu}$ and $x_{-}^{\mu}$, respectively. The common boundary of
the domains is denoted by $\sum$. In other words, the Manifolds $M_+$ and
$M_-$ are glued together along the common boundary $\sum$. The equation of
$\sum$ is written as $\phi (x^{\mu})=0$, where $\phi$ is a smooth function,
and $x^{\mu}$ is a single chart called admissible coordinate system (e.g.,
skew- Gaussian coordinates attached to geodesics) that covers the overlap
and reaches into both domains. The domains of $M$ in which $\phi$ is positive
or negative are contained in $M_{+}$ or $M_{-}$, respectively [8]. By
applying the coordinate transformations $x_{\pm}^{\mu}=x_{\pm}^{\mu}
(x^{\nu})$ on the corresponding domains, a pair of metrics
$g_{\alpha \beta}^{+}(x^{\mu})$  and $g_{\alpha \beta}^{-}(x^{\mu})$
is formed over $M_{+}$ and $M_{-}$ respectively, each suitably smooth
(say $C^{3}$). \\
The main step in the distributional approach is the definition of a hybrid
metric $g_{\alpha \beta}(x^{\mu})$ over $M$ which glues the metrics
$g_{\alpha \beta}^{+}(x^{\mu})$ and $g_{\alpha \beta}^{-}(x^{\mu})$ together
continuously on $\sum$:
\begin{eqnarray}
g_{\alpha \beta}=g_{\alpha \beta}^{+}\theta (\phi )+g_{\alpha \beta}^
{-}\theta (-\phi ),
\end{eqnarray}
where $\theta$ is the Heaviside step function and
\begin{eqnarray}
\left[ g_{\alpha \beta} (x^{\mu}) \right] =0.
\end{eqnarray}
We expect on $\sum$ the curvature and Ricci tensor to be proportional
to $\delta$ function. It follows from (1) and (2) that the first derivative
of $g_{\alpha \beta}$ is proportional to the step function.
The $\delta$ distribution can only occur in the second derivative of the
metric which enters linearly in the expressions for curvature and Ricci
tensor. So the only relevant terms in the Ricci tensor are[1]
\begin{eqnarray}
\check{R}_{\mu \nu} =\check{\Gamma}_{\mu \rho ,\nu}^{\rho} -
\check{\Gamma}_{\mu \nu , \rho}^{\rho}.
\end{eqnarray}
Using the metric in the form (1), we finally arrive at the following
expression for the components of the Ricci tensor proportional to $\delta$
distribution [1, see also 8 and 9]
\begin{eqnarray}
\check{R}_{\mu \nu}= \left( \frac{1}{2g} [g_{,\mu}] \partial_{\nu} \phi -
\left[ \Gamma_{\mu \nu}^{\rho} \right] \partial_{\rho} \phi \right)
\delta (\phi ),
\end{eqnarray}
where $g$ is the determinant of the metric and the partial derivatives are
done with respect to the coordinates $x^{\mu}$. The Einstein's equations 
for the dynamics of the singular hypersurface or thin shell $\sum$ is then 
conveniently written as
\begin{eqnarray}
\check{R}_{\mu \nu} = -8\pi G(\check{T}_{\mu \nu} -\frac{1}{2}
\check{T}g_{\mu \nu}).
\end{eqnarray}
The energy--momentum tensor of the shell $\check{T}_{\mu \nu}$, considered
as a distribution, is given by [1,8]
\begin{eqnarray}
\check{T}_{\mu \nu}=|\alpha |S_{\mu \nu} \delta (\phi ),
\end{eqnarray}
where $S_{\mu \nu}$ is the surface 4- tensor of energy-momentum of the shell,
and $\alpha$ is related to the unit normal four-vector $n^{\mu}$ of the
shell:
\begin{eqnarray}
n_{\mu}=\alpha^{-1}\partial_{\mu} \phi ,
\end{eqnarray}
with
\begin{eqnarray}
\alpha =\pm \sqrt{| g^{\nu \sigma} \partial_{\nu} \phi
\partial_{\sigma}\phi |}.
\end{eqnarray}
It is convenient to choose the negative (positive) sign in (8) for
time--(space--)like $\sum$ [1,6,8]. In this way the unit normal vector
$n_{\mu}$ is always directed from $M^{-}$ to $M^{+}$. Now, we require that
the admissible coordinates be such that the vector $n_{\mu}$ is in the
direction of increasing a space-- or time--like admissible coordinate
$x^{\mu}$ corresponding to time-- or space--like $\sum$. This is no
restriction to the choice of the admissible coordinates, as we are always
free to choose $M_+$ instead of $M_-$ or vise versa. Note that in general,
$n^{\mu}$ may point towards greater or smaller values of a space-- or
time--like coordinate $x_{\pm}^{\mu}$, which is the case in some of the
spherically symmetric examples we are going to consider. There we will see
the crucial role of these definitions and their leading to the sign function
needed to glue different manifolds. \\
In the following we restrict ourselves to timelike hypersurfaces. The case of
spacelike hypersurfaces is very similar and will be treated in the
appendix. Having a consistent definition of admissible coordinates
and the direction of $n^\mu$, we will look for a sign function similar to 
that in the Darmois--Israel method and crucial for the topological
classification of manifolds to be glued together. This can best be seen in
the case of spherically symmetric space times we are going to consider in
this paper. Assume a congruence of timelike hypersurfaces parallel
to $\sum$. Let $\bar{R}_{\pm}$ be the physical radius of the 2--D spheres
parallel to $\sum$ in $M^{\pm}$. Moving in the direction of $n^{\mu}$, each
of the $\bar{R}_{\pm}$ may either increase or decrease, depending on the
non--staticity and topology of $M^{\pm}$. Therefore, we may define the
following sign functions:
\begin{eqnarray}
\epsilon_{\pm}=sgn \left( n^{\mu} \partial_{\mu} \bar{R}_{\pm} \right)
\Bigl|_{\Sigma},
\end{eqnarray}
where $\epsilon_{\pm}$ take the values $+1$ or $-1$ accordingly as
$\bar{R}_{\pm}$ increase or decrease along the normal vector $n^{\mu}$
directed from $M_{-}$ to $M_{+}$. We will see in the next sections
that this sign functions correspond to that introduced in Darmois--Israel
method related to the sign of $K^{\theta^{\pm}}_{\theta}$[7, 8, 10]. \\
Now, the boundary $\sum$ maybe just a singular hypersurface or a
thin layer. In general, we may therefore assume for the boundary a surface
energy--momentum tensor $S^{\mu \nu}$ of a perfect fluid type given by
\begin{eqnarray}
S^{\mu \nu}= \sigma u^{\mu} u^{\nu} +w(h^{\mu \nu}-u^{\mu}u^{\nu}),
\end{eqnarray}
where $u^{\mu}$ is the unit four-velocity of any observer whose world line
lies within the shell; $\sigma$ and $\omega$ are respectively the
surface-energy density and tension measured by that observer. $h_{\mu \nu}$
denotes the induced three-metric on $\sum$ and is written as
\begin{eqnarray}
h_{\mu\nu}=g_{\mu\nu}+n_{\mu}n_{\nu}.
\end{eqnarray}           
We are now ready to apply the above formalism to obtain the junction
conditions in spherically symmetric spacetimes, bounded by a timelike shell.
The corresponding junction equations for spacelike shells will be given in
the Appendix.

\section{Junction equations in the static spherically symmetric space-times}

\hspace*{0.5cm}
Consider a spherical thin shell with a (2+1)--dimensional timelike history
$\sum$ in static spherically symmetric spacetimes on both sides described by
the Schwarzschild-de Sitter metrics given in the from
\begin{eqnarray}
ds^{2}\Bigl|_{\pm} =f_{\pm}(r_{\pm})dt^{2}_{\pm}-f_{\pm}^{-1}(r_{\pm})
dr_{\pm}^{2}-r^{2}_{\pm} d\Omega^{2},
\end{eqnarray}
with
\begin{eqnarray}
f_{\pm}(r_{\pm})=1-\frac{2Gm_{\pm}}{r_{\pm}}-\frac{\Lambda_{\pm} r_{\pm}^{2}
}{3},
\end{eqnarray}
where $m_{\pm}$ and $\Lambda_{\pm}$ are the mass parameters and the
cosmological constants associated with $M_{\pm}$. The equation of the
spherical timelike shell can be represented as
\begin{eqnarray}
\phi (x^{\mu})=r-R(t)=0,
\end{eqnarray}
where $R(t)$ is the radius of the shell as a function of the relevant
timelike coordinate $t$.\\
Now, we apply the following transformations to make the metric continuous
on the shell
\begin{eqnarray}
\begin{array}{cc} r_{+}=A(r,t),& \ \ \ \ r_{-}=C(r,t),\\ t_{+}=B(r,t),&
\ \ \ \ t_{-}=D(r,t). \end{array}
\end{eqnarray}
Carrying out the transformations and requiring the continuity of the metric
on $\sum$ according to (2), we obtain
\begin{eqnarray}
\left\{ \begin{array}{lr} U\equiv f_{+}B_{,t}^{2}-f_{+}^{-1}A_{,t}^{2}
\stackrel{\Sigma}{=}f_{-}D_{,t}^{2}-f_{-}^{-1}C_{,t}^{2},\\
 V\equiv f_{+}B_{,r}B_{,t}-f_{+}^{-1}A_{,r}A_{,t}\stackrel{\Sigma}{=}f_{-}D_{,r}D_{,t}-
 f_{-}^{-1} C_{,r}C_{,t},\\
 W\equiv f_{+}B_{,r}^{2}-f_{+}^{-1} A_{,r}^{2}\stackrel{\Sigma}{=}f_{-} D_{,r}^{2} -f_{-}^{-1}
 C_{,r}^{2},\\
 A\left( R(t),t\right) =C\left( R(t),t\right) =R(t), \end{array} \right.
\end{eqnarray}
where $\stackrel{\Sigma}{=}$ means that both sides of the equality are
evaluated on $\sum$. The sign functions $\epsilon_{+}$ and $\epsilon_{-}$
as defined by (9) take the forms
\begin{eqnarray}
\begin{array}{c} \epsilon_{+}=sgn (n^{\mu}\partial_{\mu}A)\Bigr|_{\Sigma}, \\ \\
\epsilon_{-}=sgn (n^{\mu}\partial_{\mu}C)\Bigr|_{\Sigma},\end{array}
\end{eqnarray}
where $n_{\mu}$ defined by (7) is given by
\begin{eqnarray}
n_{\mu}=|Y| (\dot{R},-\dot{t},0,0) \Bigl|_{\Sigma},
\end{eqnarray}
with
\begin{eqnarray}
Y^{2}=V^{2}-UW.
\end{eqnarray}
Using (16) and (18) we obtain finally
\begin{eqnarray}
\epsilon_{+}=\zeta_{+}sgn (f_{+}\dot{B})\Bigl|_{\Sigma},\\
\epsilon_{-}=\zeta_{-}sgn (f_{-}\dot{D}) \Bigl|_{\Sigma},
\end{eqnarray}
where $\zeta_{\pm}$ are
\begin{eqnarray}
\zeta_{+}=sgn (B_{,t}A_{,r}-B_{,r}A_{,t}) \Bigl|_{\Sigma},\\
\zeta_{-}=sgn (D_{,t}C_{,r}-D_{,r}C_{,t}) \Bigl|_{\Sigma}.
\end{eqnarray}
Comparing (20) and (21) with the signs of $K_{\theta}^{\theta \pm}$ obtained
by the Darmois-Israel method, we can see that $\zeta_{\pm}$ generated by the
transformations (15) in the distributional method, are equivalent to the
undetermined signs of the normals $n_{\mu}^{\pm}$ defined in the Darmois-
Israel approach [6,7,10]. It is assumed that $t_{\pm}$ and $\tau$, the shell
proper time, are future directed, so that $\dot{B}$ and $\dot{D}$ are
positive. Then in the $R$ region mentioned by Berezin et al [6,11], where
$f_{\pm}\Bigl|_{\Sigma}>0$, the sign factors $\zeta_{\pm}$ differenciate the
interior-exterior characters of $M_{\pm}$ [7,10], and the sign functions
$\epsilon_{\pm}$ are just determined by $\zeta_{\pm}$ according to (20)
and (21). Thus the global topology of static spherically
symmetric space-times is determined by $\epsilon_{\pm}$: if
$\epsilon_{+}=\epsilon_{-}$, then we have an ordinary centered shell
(black hole type matching); if $\epsilon_{-}=-1$ and $\epsilon_{+}=+1$, then
we have a shell in a space-time with no center (wormhole matching);if
$\epsilon_{-}=+1$ and $\epsilon_{+}=-1$, then the shell is in a space-time
with two centers (anti-wormhole matching)[10--13]. Note that these matchings 
can best be visualized in terms of Kruskal coordinates. Section I of Kruskal 
diagram corresponds to $\epsilon = +1$ and section IV corresponds to 
$\epsilon = -1$. Therefore an ordinary centered shell corresponds to matching
of two section I or two section IV manifolds. It is also possible to glue different 
part of  section I to IV or vise versa which leads to other possibilities 
discussed above[11]. \\
We restrict now, without loss of generality of the matching, the
transformation (15) on the chart $x_{-}^{\mu}$ to the conditions
\begin{eqnarray}
t_{-}=t, \ \ \ \ \ \ \ \ C_{,r}\Bigl|_{\Sigma} =\zeta_{-} .
\end{eqnarray}
This allows us to generate the sign function $\zeta$ properly. Note that in
the original distributional formalism of Mansouri-Khorrami it was set
$\zeta_- = 1$, which means no coordinate transformation of the chart
$x_{-}^{\mu}$[3, 4]. This was a restriction to the topology of the manifold
$M^-$. Using (24), the set of equations (16) can be solved for the unknown
functions in terms of $\dot{B},\dot{t}$ and $\dot{R}$:
\begin{eqnarray}
\left\{ \begin{array}{lr} C_{,t}\Bigl|_{\Sigma}=\frac{\dot{R}}{\dot{t}}(1-\zeta_{-})\Bigl|_{\Sigma},\\
\\
A_{,r}\Bigl|_{\Sigma}=-\zeta_{-}f_{-}^{-1}\dot{R}^{2}+\zeta_{+}f_{+}\dot{B}\dot{t}\Bigl|_{\Sigma},\\
\\
A_{,t}\Bigl|_{\Sigma}=\frac{\dot{R}}{\dot{t}}+\zeta_{-}f_{-}^{-1} \frac{\dot{R}^{3}}{\dot{t}}
-\zeta_{+}f_{+}\dot{B}\dot{R}\Bigl|_{\Sigma},\\
\\
B_{,r}\Bigl|_{\Sigma}=-\zeta_{-}f_{-}^{-1}\dot{B}\dot{R}+\zeta_{+}f_{+}^{-1}\dot{R}\dot{t}\Bigl|_{\Sigma},\\
\\
B_{,t}\Bigl|_{\Sigma}=\frac{\dot{B}}{\dot{t}}+\zeta_{-}\frac{\dot{B}\dot{R}^{2}}{f_{-}\dot{t}}
-\zeta_{+}f_{+}^{-1}\dot{R}^{2}\Bigl|_{\Sigma}, \end{array} \right.
\end{eqnarray}
where for a given $\zeta_{-}$, the same sign factor $\zeta_{+}=\pm 1$ takes
care of the two possible solutions for $A_{,r} , A_{,t} , B_{,r}$ and
$B_{,t}$ in (16). It is easily seen from (25) that the sign of
$(B_{,t}A_{,r}-B_{,r}A_{,t})\Bigr|_{\Sigma}$
is determined by $\zeta_{+}$ independent of the sign factor $\zeta_{-}$,
in accordance with (22). It is also seen that the metric
$g^{\pm}_{\alpha \beta}(x^{\mu})$ on $\Sigma$ takes a diagonal form
$(C_{,t}=0)$ for static shells $(\dot{R}=0)$ or when $M_{-}$ has a center
$(\zeta_{-}=+1)$.\\
We would like to note that in principle one may choose the restriction
$r_{-}=r$ and $D_{,t}\Bigl|_{\Sigma}=\zeta_{-}$ with $\dot{D}=\dot{t}$ instead
of (24). This, however, leads to
$D_{,r}\Bigl|_{\Sigma}=\frac{\dot{t}}{\dot{R}} (1-\zeta_{-})\Bigl|_{\Sigma}$
which becomes divergent for static shells $(\dot{R}=0)$. \\
To proceed further, we need the derivatives of the time coordinates on both
sides of the shell with respect to its proper time, $\tau$. It is easily seen
that
\begin{eqnarray}
\hspace*{0.5cm}\dot{B}=f_{+}^{-1} \sqrt{f_{+}+\dot{R}^{2}}\Bigl|_{\Sigma},&\hspace{1cm}
\dot{t}=f_{-}^{-1} \sqrt{f_{-}+\dot{R}^{2}}\Bigl|_{\Sigma}. 
\end{eqnarray} 
>From (4) we obtain the nonzero components of Ricci tensor: 
\begin{eqnarray}
\check{R}_{22}=R\left( \zeta_{-}\frac{\dot{R}}{\dot{t}f_{-}}A_{,t}+
f_{-}A_{,r} -\zeta_{-}f_{-}-\frac{\dot{R}^{2}}{\dot{t}^{2}f_{-}}A_{,r}
(1-\zeta_{-})\right) \Bigr|_{\Sigma} ,
\end{eqnarray}
\begin{eqnarray}
\check{R}_{00}=2\frac{\dot{R}^{2}}{R^{2}}\check{R}_{22}-\frac{U}{2\dot{R}\dot{t}}
\Bigl( f_{-,t}\dot{t}^{2}-f^{-1}_{-,t}\dot{R}^{2}-\frac{2\dot{R}}{f_{-}}
\frac{dC_{,t}}{d\tau} -f_{+,t}\dot{B}^{2}+f^{-1}_{+,t}\dot{R}^{2}-2f_{+}\dot{B}\frac{dB_{,t}}{d\tau}
\end{eqnarray}
\[+\frac{2\dot{R}}{f_{+}}
\frac{dA_{,t}}{d\tau}\Bigr) \Bigr|_{\Sigma},\]
\begin{eqnarray}
\check{R}_{10}=\check{R}_{01}=\frac{V}{U}\check{R}_{00}-\frac{2}{R^{2}}\left( \dot{R}\dot{t}
+\frac{V}{U}\dot{R}^{2}\right) \check{R}_{22},
\end{eqnarray}
\begin{eqnarray}
\check{R}_{11}=\frac{W}{U}\check{R}_{00}+\frac{2}{R^{2}}\left( \dot{t}^{2}
-\frac{W}{U}\dot{R}^{2}\right) \check{R}_{22},
\end{eqnarray}
\begin{eqnarray}
\check{R}_{33}=\sin^{2}\theta \check{R}_{22}.
\end{eqnarray}
Using the definition (9) and the unit timelike vector field $u^{\mu}$
on $\sum$,
\begin{eqnarray}
u^{\mu}=(\dot{t},\dot{R},0,0),
\end{eqnarray}
we obtain $\check{T}_{\mu\nu}$ for the nondiagonal metric
$g_{\mu\nu}^{\pm} (x^{\sigma})$ on the shell:
{\large
\begin{eqnarray}
\check{T}_{\mu\nu}=\left[ \begin{array} {cccc}
\frac{\sigma (U\dot{t}+V\dot{R})^{2}}{\dot{t}}&\frac{\sigma (U\dot{t}+V\dot{R})(V\dot{t}+W\dot{R})}{\dot{t}}
&0&0\\ \frac{\sigma (U\dot{t}+V\dot{R})(V\dot{t}+W\dot{R})}{\dot{t}}&
\frac{\sigma (V\dot{t}+W\dot{R})^{2}}{\dot{t}}&0&0\\ 0&0&-\frac{\omega R^{2}}{\dot{t}}
&0\\ 0&0&0&-\frac{wR^{2}\sin^{2}\theta}{\dot{t}} \end{array}
\right]_{\Bigl|_{\Sigma}}.
\end{eqnarray}
}
The junction equations and dynamics of the singular hypersurface is given by
the equation (5). In our case, this reduces to the following two independent
equations:
\begin{eqnarray}
\check{R}_{22}=-8\pi G\check{P}_{22}, \hspace{2cm}\\
\check{R}_{00}-\frac{2\dot{R}^{2}}{r^{2}_{-}}\Bigl|_{\Sigma}\check{R}_{22}
=-8\pi G \left( \check{P}_{00}-\frac{2\dot{R}^{2}}{r^{2}_{-}}\Bigl|_{\Sigma}
\check{P}_{22} \right),
\end{eqnarray}
where $\check{P}_{\mu \nu}$ denotes the right hand side of (5):
$$\check{P}_{\mu \nu}=\check{T}_{\mu \nu}-\frac{1}{2}\check{T}g_{\mu \nu}.$$
These are equivalent to the set of equations (2.59a) and (2.59b) in Ref. [6].
Making use of relations obtained so far, we obtain after some manipulations
the junction equations in the final form:
\begin{eqnarray}
\epsilon_{-}\sqrt{f_{-}+\dot{R}^{2}}-\epsilon_{+}\sqrt{f_{+}+\dot{R}^{2}}
\stackrel{\Sigma}{=}4\pi G\sigma R, \hspace{1.3cm}\\
\frac{d}{d\tau}\left( \epsilon_{+}\sqrt{f_{+}+\dot{R}^{2}}-\epsilon_{-}
\sqrt{f_{-}+\dot{R}^{2}}\right) \stackrel{\Sigma}{=}8\pi G\dot{R} \left(
\frac{\sigma}{2}-w\right),
\end{eqnarray}
where we have inserted the sign functions $\epsilon_{\pm}$ instead of
$\zeta_{\pm}$ according to (20) and (21) for static spherically symmetric
spacetimes. These junction equations are equivalent to those obtained by
Berezin et al [6].

\section{Junction equations for FRW space-times}

\hspace*{0.5cm}
Consider a spherical thin shell with timelike history $\sum$ as the boundary
of two Friedmann-Robertson-Walker spacetimes with metrics given by
\begin{eqnarray}
ds^{2}\Bigl|_{\pm}=dt^{2}_{\pm}-a^{2}_{\pm}(t_{\pm})\left[ d\chi^{2}_{\pm}
+r^{2}_{\pm}(\chi_{\pm} )d\Omega^{2}\right] ,
\end{eqnarray}
where
\begin{eqnarray}
r(\chi )=\left\{ \begin{array}{lr} \sin \chi \hspace*{1.2cm}
(k=+1, \ {\rm closed \ universe}),\\
\chi \hspace{1.8cm}(k=0 , \ {\rm flat \ universe}),\\
\sinh \chi \hspace{1cm} (k=-1, \ {\rm open \ universe}). \end{array}
\right.
\end{eqnarray}
We now apply the following transformations to make the four--metric continuous
on $\sum$:
\begin{eqnarray}
\begin{array}{cc} \chi_{+}=A(r,t),& \ \ \ \ \ \ \chi_{-}=C(r,t),\\
t_{+}=B(r,t),&\ \ \ \ \ \ t_{-}=D(r,t). \end{array}
\end{eqnarray}
According to (2), we get
\begin{eqnarray}
\left\{ \begin{array}{lr} B^{2}_{,t}-a_{+}^{2}A_{,t}^{2}\stackrel{\Sigma}{=}
D^{2}_{,t}-a_{-}^{2}C^{2}_{,t},\\
B_{,r}B_{,t}-a^{2}_{+}A_{,r}A_{,t}\stackrel{\Sigma}{=}D_{,r}
D_{,t}-a^{2}_{-}C_{,r}C_{,t},\\
B^{2}_{,r}-a^{2}_{+}A^{2}_{,r}\stackrel{\Sigma}{=}D^{2}_{,r}-a^{2}_{-}C^{2}_{,r},\\
l\equiv a_{+}(B(R(t),t))r_{+}(A(R(t),t))=a_{-} (D(R(t),t))r_{-} (C(R(t),t)),
\end{array} \right.
\end{eqnarray}
where $\phi (x^{\mu})=r-R(t)=0$ is the equation of the shell and $l$
represents the physical radius of it.\\
The sign functions $\epsilon_{+}$ and $\epsilon_{-}$ as defined by (9) are
given by 
\begin{eqnarray}
\begin{array}{c} \epsilon_{+}=sgn \left( n^{\mu}\partial_{\mu} (a_{+}r_{+}) \right)
|_{\Sigma}, \\
\epsilon_{-}=sgn \left( n^{\mu}\partial_{\mu} (a_{-}r_{-})\right) |_{\Sigma}.
\end{array}
\end{eqnarray}
Using $n_{\mu}$ as obtained in (18) we obtain after some manipulations
\begin{eqnarray}
\epsilon_{+}=\zeta_{+}sgn \left( \frac{dr_{+}}{d\chi_{+}}+lH_{+} \frac{a_{+}
\dot{A}}{\dot{B}}\right) \Bigl|_{\Sigma}, \\
\epsilon_{-}=\zeta_{-}sgn \left( \frac{dr_{-}}{d\chi_{-}}+lH_{-} \frac{a_{-}
\dot{C}}{\dot{D}}\right) \Bigr|_{\Sigma},
\end{eqnarray}
where $H_{\pm}$ are the Hubble parameters for $M_{\pm}$, and the sign factors
$\zeta_{+}$ and $\zeta_{-}$ are the same as that given by (22) and (23)
respectively, differentiating the interior-exterior characters of a flat or
open universe(for a closed universe one can change the signs of $\zeta_{\pm}$
by the coordinate transformations $\chi_{\pm} \longrightarrow \pi -\chi_
{\pm}$, see Ref.[10]). Note that these relations correspond to Eq.(5) in
Ref. [7]. \\
Similarly, we have assumed $t_{\pm}$ and $\tau$ to be future directed. For
given $\zeta_{\pm}$ and $k_{\pm}$, one can determine the topology of FRW
space times regardless of the signs of $\epsilon_{\pm}$. (A list of possible
topology types has been given in Ref. [7]). In other words, we may have any
sign of $\epsilon_{\pm}$ for any known value of $\zeta_{\pm}$ and $k_{\pm}$.
Particularly for $k_{+}\leq 0$ and $\zeta_{\pm} >0$, the sign of $\epsilon_{+}$
can be negative, if the comoving radius of the shell decreases in time, namely
if $\dot{A}<0$ or equivalently, if the peculiar velocity of the shell
$v^{+}_{pe} = a_{+}\frac{\dot{A}}{\dot{B}} \Bigl|_{\Sigma}$ observed in
$M_{+}$ is negative. Therefore, when the second term in the expression within the
brackets in (43), which is negative, becomes large, $\epsilon_{+}$ becomes
negative. It can be shown that in this case the size of the shell is larger
than the Hubble length in $M_{+}$, i.e. $H_{+}^{-1}$, as mentioned by Sakai
and Maeda [7]. The possibility of negative value for $\epsilon_{+}$ in the
flat or open universe $(k_{+}\leq 0)$ had been missed in Ref. [6].\\
Again, without loss of generality, we may restrict the transformations(40) on
the chart $x_{-}^{\mu}$ by requiring
\begin{eqnarray}
t_{-}=t, \hspace*{2cm}C_{,r}\Bigl|_{\Sigma}=\zeta_{-},\hspace*{2cm}
C_{,t}\Bigl|_{\Sigma} = 0.
\end{eqnarray}
Using these restrictions we obtain
\begin{eqnarray}
\left\{ \begin{array}{lr} \dot{R}=\zeta_{-}\dot{C}\Bigl|_{\Sigma},\\
\\
A_{,r}\Bigl|_{\Sigma}=-\zeta_{-}a_{-}^{2}\dot{C}\dot{A}+\zeta_{+}\frac{a_{-}}{a_{+}} \dot{B}
\dot{t}\Bigl|_{\Sigma}, \\
\\
A_{,t}\Bigl|_{\Sigma}=\dot{A}\dot{t}-\zeta_{-}\zeta_{+}\frac{a_{-}}{a_{+}} \dot{B}
\dot{C}\Bigl|_{\Sigma},\\
\\
B_{,r}\Bigl|_{\Sigma}=-\zeta_{-}a_{-}^{2}\dot{B}\dot{C}+\zeta_{+}a_{-}a_{+} \dot{A}\dot{t}\Bigl|_{\Sigma},\\
\\
B_{,t}\Bigl|_{\Sigma}=\dot{B}\dot{t}-\zeta_{-}\zeta_{+}a_{-}a_{+} \dot{A}
\dot{C}\Bigl|_{\Sigma},
\end{array} \right.
\end{eqnarray}
where for given $\zeta_{-}$, the same sign factor $\zeta_{+}=\pm 1$ takes
care of the two possible solutions for $A_{,r} , A_{,t} , B_{,r}$ and
$B_{,t}$ in (41). Remarkably, from (46) we may show that sign of
$(B_{,t}A_{,r}-B_{,r}A_{,t}))\Bigl|_{\Sigma}$
in accordance with (22) is fixed by $\zeta_{+}$ regardless of $\zeta_{-}$. \\
We were free to choose instead of the restriction (45) the transformation\\
$$t_{-}=t,\hspace*{2cm} C_{,r}\Bigl|_{\Sigma}=\zeta_{-}, \hspace*{2cm} \dot{C}=\dot{R}$$. 
This would have the disadvantage of leading to a nondiagonal continuous
metric on $\Sigma$. \\
Following relations between $B$, $A$, $C$, $t$, and $l$ is obtained by the
requirement that $\tau$ is the proper time on $\sum$ seen from $M_{-}$ or
$M_{+}$:
\begin{eqnarray}
\dot{B}^{2}-a_{+}^{2}\dot{A}^{2}\Bigl|_{\Sigma}=1, \hspace*{4cm} \dot{t}^{2}
- a_{-}^{2}\dot{C}^{2}\Bigl|_{\Sigma}=1.
\end{eqnarray}
Differentiating the fourth equation in(41) with respect to the shell
proper time $\tau$ leads to 
\begin{eqnarray}
\dot{l}=a_{-} \frac{dr_{-}}{d\chi_{-}}\dot{C} +lH_{-}
\dot{t}\Bigl|_{\Sigma}, \ \ \ \ \ \ \ \ \ \
\dot{l}=a_{+} \frac{dr_{+}}{d\chi_{+}}\dot{A}+lH_{+} \dot{B}\Bigl|_{\Sigma}.
\end{eqnarray}
Solving (47) and (48) for $\dot{B}$ and $\dot{t}$, we obtain
\begin{eqnarray*}
\dot{B}=\frac{-l\dot{l}H_{+}+\sqrt{(1+\dot{l}^{2} -\frac{8\pi G}{3}
\rho_{+}l^{2})(1+l^{2}H_{+}^{2}-\frac{8\pi G}{3}\rho_{+}l^{2})}}{1-\frac{8\pi G}{3}
\rho_{+}l^{2}}\Bigl|_{\Sigma},
\end{eqnarray*}
\begin{eqnarray}
\dot{t}=\frac{-l\dot{l}H_{-}+\sqrt{(1+\dot{l}^{2} -\frac{8\pi G}{3} \rho_{-}
l^{2})(1+l^{2}H_{-}^{2}-\frac{8\pi G}{3}\rho_{-}l^{2})}}{1-\frac{8\pi G}{3}
\rho_{-} l^{2}} \Bigl|_{\Sigma},
\end{eqnarray}
where we have used the following Friedmann equation for $M_{\pm}$:
\begin{eqnarray}
H_{\pm}^{2}+\frac{k_{\pm}}{a^{2}_{\pm}}=\frac{8\pi G}{3}\rho_{\pm},
\end{eqnarray}
with $\rho_{\pm}(t_{\pm})$ being the energy density in $M_{\pm}\cdot \dot{B}$
and $\dot{t}$ are positive in $R$ or $T$ regions where the denominators
in (49) are positive or negative, respectively [6].\\
The nonzero components of Ricci tensor computed from (4) are
\begin{eqnarray}
\check{R}_{00}=\frac{2\dot{R}^{2}a^{2}_{-}}{l^{2}} \check{R}_{22} +\frac{1}{a^{2}_{-}
\dot{R}\dot{t}} \left( \dot{B} \frac{dB_{,t}}{d\tau} -a_{+}^{2}\dot{A}
\frac{dA_{,t}}{d\tau}-a_{+}^{2}H_{+}\dot{A}^{2}B_{,t}+a_{-}^{2}
\dot{R}^{2} H_{-}\right) \Bigr|_{\Sigma},
\end{eqnarray}
\begin{eqnarray}
\check{R}_{10}=\check{R}_{01}=\frac{-2a_{-}^{2}\dot{R}\dot{t}}{l^{2}} \check{R}_{22} \Bigl|_{\Sigma},
\end{eqnarray}
\begin{eqnarray}
\check{R}_{11}=-a_{-}^{2}\check{R}_{00}+\frac{2a^{2}_{-}}{l^{2}} (\dot{t}^{2}+a_{-}^{2}
\dot{R}^{2}) \check{R}_{22} \Bigl|_{\Sigma},
\end{eqnarray}
\begin{eqnarray}
\check{R}_{22}=\frac{l^{2}}{\dot{R}\dot{t}a^{2}_{-}} \left( H_{-}-
H_{+} B_{,t}-\frac{a_{+}}{l}\frac{dr_{+}}{d\chi_{+}}A_{,t}\right)  \Bigl|_{\Sigma},
\end{eqnarray}
\begin{eqnarray}
\check{R}_{33}=\sin^{2}\theta \check{R}_{22}.
\end{eqnarray}
Likewise, we obtain $\check{T}_{\mu\nu}$ for the diagonal metric
$g_{\mu\nu}^{\pm} (x^{\sigma})$ on $\sum$:
\begin{eqnarray}
\check{T}_{\mu\nu} =\frac{1}{\dot{t}a_{-}} \left[ \begin{array}{cccc}
\sigma \dot{t}^{2}&-\sigma a^{2}_{-}\dot{R}\dot{t}&0&0\\ -\sigma a^{2}_{-}
\dot{R}\dot{t}&\sigma a_{-}^{4}\dot{R}^{2}&0&0\\ 0&0&-wl^{2}&0\\
0&0&0&-wl^{2}\sin^{2}\theta \end{array} \right]_{\Bigl|_{\Sigma}}.
\end{eqnarray}
Here again we are faced with two independent Einstein equations (34) and (35)
on the shell. The junction equations of FRW space-times are then obtained
after some manipulations: 
\begin{eqnarray}
\zeta_{-}\dot{t} \left( \frac{dr_{-}}{d\chi_{-}}+lH_{-} \frac{a_{-}
\dot{C}}{\dot{t}}\right) -\zeta_{+}\dot{B} \left( \frac{dr_{+}}{d\chi_{+}}+
lH_{+} \frac{a_{+}\dot{A}}{\dot{B}}\right) \stackrel{\Sigma}{=}4\pi Gl\sigma ,\\
\zeta_{+}\left( \frac{\ddot{B}}{a_{+}\dot{A}}+a_{+}\dot{A}H_{+}\right) -
\zeta_{-} \left( \frac{\ddot{t}}{a_{-}\dot{C}}+a_{-}\dot{C}H_{-}\right)
\stackrel{\Sigma}{=}8\pi G(\frac{\sigma}{2}-w).
\end{eqnarray}
In the framework of Darmois-Israel method, the left hand sides of (57) and
(58) show the jump of extrinsic curvatures $K_{\theta}^{\theta}$ and
$K_{\tau}^{\tau}$ across the shell, respectively [7,10].  \\
Now, using (43) and (44) for the sign functions $\epsilon_{\pm}$, the
junction equation (57) can be written in the following final from:
\begin{eqnarray}
\epsilon_{-} \sqrt{1+\dot{l}^{2}-\frac{8\pi G}{3}\rho_{-}l^{2}}
-\epsilon_{+}\sqrt{1+\dot{l}^{2}-\frac{8\pi G}{3} \rho_{+}l^{2}}
\stackrel{\Sigma}{=}4\pi Gl\sigma ,
\end{eqnarray}
which is the same as that obtained by Berezin et al [6]. To eliminate the
metric functions from the second junction equation, we apply the second
Friedmann equation: 
\begin{eqnarray}
\frac{1}{a_{\pm}} \frac{da^{2}_{\pm}}{dt^{2}_{\pm}}=-\frac{4\pi G}{3}
\left( \rho_{\pm}+3p_{\pm}\right),
\end{eqnarray}
where $p_{\pm}$ denote the pressures of the matter in $M_{\pm}$.
Finally, after some lengthy manipulations, the equation (58) can be written
as follows
\[\frac{\zeta_{+}}{\sqrt{1+\dot{l}^{2}-\frac{8\pi G}{3}\rho_{+}l^{2}}}
\Bigl\{ \ddot{l}-\frac{8\pi G}{3}\rho_{+}l+4\pi Gl(\rho_{+}+p_{+}) \]
\[\cdot \left( 1+\frac{2l^{2}\dot{l}^{2}H_{+}^{2}}{\Delta_{+}^{2}} +\frac{1}{\Delta_{+}}
\left( \dot{l}^{2}+l^{2}H_{+}^{2}-\frac{2l\dot{l}H_{+}}{\Delta_{+}}
\sqrt{l^{2}\dot{l}^{2}H_{+}^{2}+\Delta_{+}^{2}+\Delta_{+} (\dot{l}^{2}+l^{2}H_{+}^{2}
)} \right) \right) \Bigr\} \]
\[-\frac{\zeta_{-}}{\sqrt{1+\dot{l}^{2}-\frac{8\pi G}{3}\rho_{-}l^{2}}} \Bigl\{
\ddot{l} -\frac{8\pi G}{3}\rho_{-}l+4\pi Gl(\rho_{-}+p_{-})\]
\[\cdot \left( 1+\frac{2l^{2}\dot{l}^{2}H_{-}^{2}}{\Delta_{-}^{2}}+\frac{1}{\Delta_{-}}
\left( \dot{l}^{2}+l^{2}H_{-}^{2}-\frac{2l\dot{l}H_{-}}{\Delta_{-}} \sqrt{
l^{2}\dot{l}^{2}H_{-}^{2}+\Delta_{-}^{2}+\Delta_{-} (\dot{l}^{2} +l^{2}H_{-}^{2})}
\right)\right) \Bigr\} \]
\begin{eqnarray}
\stackrel{\Sigma}{=}8\pi G(\frac{\sigma}{2}-w),
\end{eqnarray}
where $\Delta_{\pm}$ is given by 
\begin{eqnarray}
\Delta_{\pm}=1-\frac{8\pi G}{3} \rho_{\pm}l^{2} \Bigl|_{\Sigma}.
\end{eqnarray}
The same result has been obtained by Berezin et al [6].

\section{Conclusion}

\hspace*{0.5cm}
In glueing different manifolds together, the global characterization of the
glued manifold is of utmost importance. This characterization depends
basically on the signs of $K^{\theta^{\pm}}_{\theta}$, which is the central
quantity in the Darmois--Israel approach. So far such global classifications
had been neglected within the distributional approach of Mansouri--Khorrami.
Now, we have been able to define the relevant sign functions
$\epsilon_{\pm}$ properly by the equation (9), and to insert them in the
junction equations, so that the global classification of spherically
symmetric space-times glued together can also be done within this
distributional formalism. This makes the Mansouri--Khorrami formalism as
general as possible and, in some cases, more suitable to apply than
Darmois--Israel method. It has been successfully applied to solve special
problems like the dynamics of two shells[14] or thick shells [15] within
general relativity \\
We have also seen, given the unique normal $n_{\mu}$ in the
distributional formalism, that the sign factors  $\zeta_{+}$ and
$\zeta_{-}$ are independently generated by applying the transformations
$x_{+}^{\mu}=x_{+}^{\mu} (x^{\nu})$ and $x_{-}^{\mu}=x_{-}^{\mu} (x^{\nu})$
on $M_{+}$ and $M_{-}$ respectively. This is in contrast to the
Darmois--Israel method in which the sign factors are inserted in front of the
normal vectors $n_{\mu}^{\pm}$. Our sign factors are naturally related to
the interior-exterior characterizations of $M_{\pm}$ [7,10]. In addition, we
have seen that to include the most general case of glueing manifolds in the
distributional formalism along the singular hypersurface it is not enough
to make a coordinate transformation on just one of the manifolds to make the
four--metric continuous. We have to apply the nontrivial transformations
$x_{\pm}^{\mu} = x_{\pm}^{\mu} (x^{\nu})$ on both sides of the shell to
generate the sign factors $\zeta_{\pm}$. These generalizations had been
neglected in the papers [3,4]. It should also be noted that admissible
coordinates have to be such that the relevant coordinate along the vector
$n^{\mu}$ is increasing.

\section*{Appendix: Junction equations in the spherically symmetric
space--times separated by a spacelike shell}

\hspace{0.5cm}
The following junction equations for FRW space times bounded by a space--like
shell can be obtained from (59) and (61) by means of the substitutions
$\tau \longrightarrow i\xi$, $\sigma \longrightarrow -i\sigma$ and
$\omega  \longrightarrow -i\omega$, where $\xi$ is the distance from the
center of the sphere of the radius $R(\xi )$:
\[\hspace*{2cm}\epsilon_{-} \sqrt{l'^{2}-1+\frac{8\pi G}{3} \rho_{-}l^{2}} 
-\epsilon_{+}\sqrt{l'^{2}-1+\frac{8\pi G}{3}\rho_{+}l^{2}}\stackrel{\Sigma}
{=}-4\pi Gl\sigma , \hspace*{2cm} (A1)\]
\\
\[\frac{\zeta_{+}}{\sqrt{l'^{2}-1+\frac{8\pi G}{3}\rho_{+}l^{2}}}
\Bigl\{ -l'' -\frac{8\pi G}{3}\rho_{+}l+4\pi Gl(\rho_{+}+p_{+}) \]
\[\cdot \left( 1-\frac{2l^{2}l'^{2}H_{+}^{2}}{\Delta_{+}^{2}} +\frac{1}{\Delta_{+}}
\left( -l'^{2}+l^{2}H_{+}^{2}-\frac{2ll' H_{+}}{\Delta_{+}}
\sqrt{l^{2}l'^{2}H_{+}^{2}-\Delta_{+}^{2}-\Delta_{+} (-l'^{2}+l^{2}H_{+}^{2}
)} \right) \right) \Bigr\} \]
\[-\frac{\zeta_{-}}{\sqrt{l'^{2}-1+\frac{8\pi G}{3}\rho_{-}l^{2}}} \Bigl\{
-l'' -\frac{8\pi G}{3}\rho_{-}l+4\pi Gl(\rho_{-}+p_{-})\]
\[\cdot \left( 1-\frac{2l^{2}l'^{2}H_{-}^{2}}{\Delta_{-}^{2}}+\frac{1}{\Delta_{-}}
\left( -l'^{2}+l^{2}H_{-}^{2}-\frac{2ll'H_{-}}{\Delta_{-}}
\sqrt{l^{2}l'^{2}H_{-}^{2}-\Delta_{-}^{2}-\Delta_{-} (-l'^{2} +l^{2}H_{-}^{2})}
\right)\right) \Bigr\} \]
\[\hspace*{5cm}\stackrel{\Sigma}{=}-8\pi G(\frac{\sigma}{2}-w), \hspace{5.5cm} (A2)\]
\[{\rm where} \ l'\equiv \frac{dl}{d\xi}.\hspace*{12.8cm}\]

\section*{References}
\hspace{0.6cm}[1] R. Mansouri and M. Khorrami, J. Math. Phys. 37,5672 (1996).

[2] A. Lichnerowicz, in Relativity, Quanta, and Cosmology (Einstein 1879-1979), vol \\
\hspace*{0.6cm}II Johnson, New York (1979).

[3] M. Khorrami and R. Mansouri, J. Math. Phys. 35,951 (1994).

[4] M. Khorrami and R. Mansouri, Phys. Rev. D44, 557 (1991).
      
[5] H. Sato, Prog. Theor. Phys. 76,1250 (1986).

[6] V. A. Berezin, V. A. Kuzmin, I. I. Tkachev, Phys. Rev. D36, 2919 (1987).

[7] N. Sakai and K. Maeda, Phys. Rev. D50, 5425 (1994).

[8] C. Barrabes and W. Israel, Phys. Rev. D43, 1129 (1991).

[9] M. Mars and J. M. M. Senovilla, Class. Quantum Grav. 10, 1865 (1993).

[10] N. Sakai and K. Maeda, Prog. Theor. Phys. 90, 1001 (1993).

[11] V. A. Berezin, A. M. Boyarsky and A. Yu. Neronov, Phys. Rev. D57, 1118\\
\hspace*{0.7cm}(1998).

[12] D. S. Goldwirth and J. Katz, Class. Quantum Grav. 12,769 (1995).

[13] P. Laguna-Castillo and R. A. Matzner, Phys. Rev. D34, 2913 (1986).

[14] S. Khakshournia and R. Mansouri, dynamics of two spherically symmetric shells\\
\hspace*{0.7cm} within general relativity, paper in preparartion.

[15] S. Khakshournia and R. Mansouri, dynamics of thick spherically symmetric shells\\
\hspace*{0.7cm} within general relativity, paper in preparartion.

\end{document}